\documentclass{epl}
\usepackage{graphicx}
\usepackage{bm}

\title{Small scale statistics of viscoelastic turbulence}
\shorttitle{Small scale viscoelastic turbulence}
\author{S. Berti \inst{1}, A. Bistagnino\inst{1},
G. Boffetta\inst{1}, A. Celani\inst{2} and S. Musacchio\inst{2}}
\institute{
  \inst{1} Dipartimento di Fisica Generale and INFN,
           Universit\`a degli Studi di Torino,
           Via Pietro Giuria 1, 10125, Torino, Italy \\
           and ISAC-CNR, Sezione di Torino, Italy \\
  \inst{2} INLN-CNRS, 1361 route des Lucioles, Sophia Antipolis, 06560 Valbonne, France.
}
\pacs{47.27.Gs}{Isotropic turbulence; homogeneous turbulence}
\pacs{47.27.E-}{Turbulence simulation and modeling}
\pacs{47.57.Ng}{Polymers and polymer solutions}

\begin{document}

\maketitle
\begin{abstract}
The small scale statistics of homogeneous isotropic turbulence of 
dilute polymer solutions is investigated by means of direct numerical 
simulations of a simplified viscoelastic fluid model.
It is found that polymers only partially suppress the turbulent cascade 
below the Lumley scale, leaving a remnant energy flux even for large 
elasticity. As a consequence, fluid acceleration in viscoelastic flows
is reduced with respect to Newtonian turbulence, whereas its rescaled 
probability density is left unchanged.
At large scales the velocity field is found to be unaffected by 
the presence of polymers.
\end{abstract}

The addition of small amounts of long chain
polymers produces dramatic effects on flowing fluids, the most
renowned being the reduction of friction drag in turbulent
flow at high Reynolds numbers \cite{GB95,SW00}.
Most studies focused on dilute polymer solutions
in channel or pipe geometry, where boundary effects are important \cite{V75},
but recent experimental \cite{AK02,LGLKT05} and numerical 
\cite{BCM03,DCBP05,BCM05}
works have shown that polymers affect the turbulent flow even far from
(or in absence of) boundaries.
In particular, Ref.\cite{DCBP05} studied the modification of
the turbulent cascade induced by polymers in numerical simulations
of homogeneous, isotropic turbulence.
In this paper we investigate the effects of polymer addition
to the small scale statistics in fully developed homogeneous-isotropic 
turbulence by means of direct numerical simulations of a simplified 
viscoelastic model.
We show that, by increasing the elasticity of polymers,
the energy flux in the turbulent cascade is partially suppressed
and transferred to the elastic degrees of freedom.
This suppression remains partial even for large values of elasticity:
as a consequence the energy flux to small scales remains finite
and the small scale statistics, such as acceleration probability
density function (pdf), retain some
characteristics of Newtonian flows.


The mechanism by which dilute polymer solutions can influence
turbulent flows is the extreme extensibility of polymers.
Polymers, typically composed by a large number of monomers,
at equilibrium are coiled in a ball of radius $R_0$.
In presence of a nonhomogeneous flow, the molecule is deformed
in an elongated structure characterized by its
end-to-end distance $R$ which can be significantly larger than $R_0$.
The deformation of molecules is the result of the competition
between the stretching induced by differences of velocities and
the entropic relaxation of polymers to their equilibrium configuration.
Experiments with DNA molecules \cite{PSC94-QBC97} show that
this relaxation is linear, provided that the elongation is
small compared with the maximal extension $R \ll R_{max}$,
and can be characterized by a typical relaxation time $\tau$ \cite{H77}.

These ingredients lead to the simplest model which describes the
behavior of a polymer in a flow, the dumbbell model.
Since in applications the typical size of polymers is smaller
than the viscous scale of turbulence, stretching is due to velocity
gradients and the end-to-end distance evolves according to:
\begin{equation}
{d {\bm R} \over d t} = ({\bm \nabla} {\bm u})^{T} {\bm R}
- {1 \over \tau} {\bm R} +
\sqrt{{2 R_0^2 \over \tau}} {\bm \xi}
\label{eq:1}
\end{equation}
where ${\bm \xi}$ is a Brownian process with correlation
$\langle \xi_i(t) \xi_j(t') \rangle=\delta_{ij} \delta(t-t')$.

The relative importance between polymer relaxation and stretching
is measured by the Weissenberg number $Wi$, defined as the product of
$\tau$ and the characteristic velocity gradient.
When $Wi \ll 1$ relaxation is fast compared to the stretching time
and polymers remain in the coiled state. For $Wi \gg 1$, on the
contrary, polymers are substantially elongated. The transition point
is called the coil-stretch transition and occurs at $Wi=O(1)$.

In the case of dilute solutions, for which the polymer concentration $n$
satisfies $n R_0^3 \ll 1$, the influence of polymers in the coiled
state on the fluid is negligible.
Above the coil-stretch transition, polymers start to affect the flow.
This regime is characterized by large elongations
$R \gg R_0$, which allow to disregard the thermal noise in (\ref{eq:1}).
Polymer solutions at macroscopic scales,
i.e. at scales much larger than typical interpolymer distances,
can be described by a local elongation field ${\bm R}({\bm x},t)$
which evolves according to
\begin{equation}
{\partial {\bm R} \over \partial t} + {\bm u} \cdot {\bm \nabla} {\bm R}
= ({\bm \nabla} {\bm u})^{T} \cdot {\bm R} - {{\bm R} \over \tau}
\label{eq:2}
\end{equation}
Taking the divergence of (\ref{eq:2}) one easily sees that
${\bm \nabla} \cdot {\bm R}$ decays in time and thus we can take
${\bm R}$ solenoidal.

The effect of polymers on the fluid is in the modification of the stress
tensor through an additional elastic component ${\bm \Pi}^{P}$ which takes
into account the elastic forces of polymers
${\bm \Pi}^{P}_{ij}=(2 \nu \eta/\tau) (R_i R_j/R_0^2)$
where $\nu$ is the solvent viscosity and $\eta$ (proportional to
polymer concentration) represents the zero-shear contribution of polymers
to the total solution viscosity $\nu (1 + \eta)$.
The Navier-Stokes equation for the incompressible velocity field
${\bm u}({\bm x},t)$ thus becomes
\begin{equation}
{\partial {\bm u} \over \partial t} + {\bm u} \cdot {\bm \nabla} {\bm u}=
- {\bm \nabla} p + \nu \triangle {\bm u} + {2 \nu \eta \over \tau} \,
{{\bm R} \cdot {\bm \nabla} {\bm R} \over R_0^2}
\label{eq:4}
\end{equation}
Equations (\ref{eq:2}) and (\ref{eq:4}) are the so-called
uniaxial model for viscoelastic flows which can be also obtained
starting from the linear Oldroyd-B model \cite{BCAH87} by
taking the limit of large elongations \cite{FL03,BSS06}.
Observe that by introducing the rescaled variable
${\bm B}=\sqrt{2 \nu \eta /\tau} ({\bm R}/R_0)$, 
equations (\ref{eq:2}-\ref{eq:4})
formally become the MHD equations for a plasma at zero resistivity
with a linear damping $-{\bm B}/\tau$ \cite{Biskamp,OP03}. In this
representation the coefficient $\eta$ disappears and thus the
dynamics of the uniaxial model is independent of the concentration
(which is physically consistent with the assumption of linearity
and strong elongation).
In the following we will thus take $\eta=1$ and absorb $R_0$ in the 
definition of $R$: $R \rightarrow R/R_0$.

The total energy (kinetic plus elastic) 
$E_T=(\langle u^2 \rangle + \langle B^2 \rangle)/2$
of the flow is dissipated at a rate
\begin{equation}
{d E_T \over d t} = 
- \nu \langle {|{\bm \nabla} \times {\bm u}|}^2 \rangle
- {1 \over \tau} \langle B^2 \rangle=
-\varepsilon_{\nu}-\varepsilon_{\tau}
\label{eq:5}
\end{equation}
where $\varepsilon_{\nu}$ is the viscous dissipation 
while the second term $\varepsilon_{\tau}$ represents the additional 
dissipation due to the relaxation of polymers to their
equilibrium configuration. 

In the following we will consider the statistics of stationary
turbulent solutions of the viscoelastic model (\ref{eq:2}-\ref{eq:4}) 
integrated in a periodic box 
of size $L=2 \pi$ at resolution $128^3$ by means of a standard
pseudo-spectral code for different values of the relaxation time $\tau$. 
For each $\tau$, a statistically stationary state
is obtained by adding to (\ref{eq:4}) an external forcing term 
which acts on the largest scales by keeping their energy 
constant \cite{CDKS93}. In stationary conditions the forcing injects 
energy with a mean rate $\varepsilon_{I}$ which balances the dissipation
(\ref{eq:5}), $\varepsilon_{I}=\varepsilon_{\nu}+\varepsilon_{\tau}$.
The turbulent regime for a viscoelastic flow is controlled 
by two dimensionless parameters, the Reynolds number 
$R_{\lambda}=u_{rms} \lambda/\nu$ (where 
$\lambda=u_{rms}/\langle (\partial_x u_x)^2\rangle^{1/2}$
is the Taylor microscale)
and the Weissenberg number which is defined here as 
$Wi=\tau/\tau_{K}$, where
$\tau_{K}=(\nu/\varepsilon_{\nu})^{1/2}$ is the Kolmogorov time.
As a reference run, we integrated the standard Navier-Stokes
equations (\ref{eq:4}) with $\eta=0$, for which we have 
$R_{\lambda} \simeq 87$.
In this Newtonian limit, we have also computed the Lagrangian 
Lyapunov exponent 
$\lambda_{L}$ which is a measure of the mean stretching rate.
The dimensionless number $\lambda_L \tau_{\eta} \simeq 0.13$
is consistent with known simulations \cite{GP90}.
The viscoelastic runs are performed for different values of
the relaxation times correspoding to a Weissenberg number in the
range $4.8 \le Wi \le 24$ (i.e. $0.63 \le \lambda_L \tau \le 3.14$).

\begin{figure}[htb]
\includegraphics[draft=false,scale=1.0]{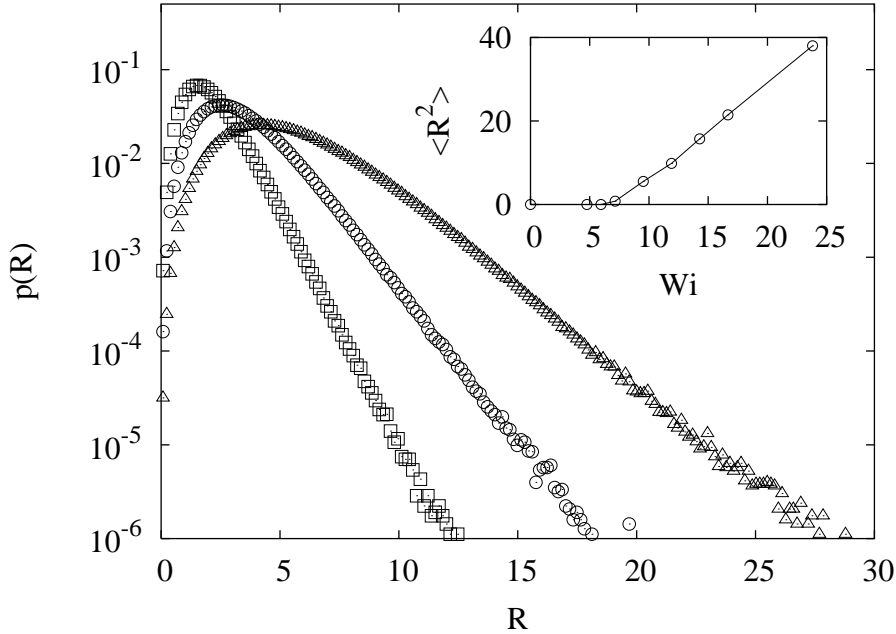}
\caption{Probability density functions of polymer elongations
above the coil-stretch transition 
at $Wi=9.5$ (squares), $Wi=12$ (circles) and $Wi=24$ (triangles).
For $Wi<Wi^{*}=6.5$ the distribution is $p(R)=\delta(R)$.
In the inset we show the growth of $\langle R^2 \rangle$ vs $Wi$.}
\label{fig0}
\end{figure}

Although the uniaxial model is derived in the limit of strong 
elongation, it displays a clear coil-stretch transition as
a function of $Wi$ \cite{L69-L73}.
The limit $Wi \to 0$, i.e. $\tau \to 0$, in (\ref{eq:2})
implies ${\bm R}=0$ and (\ref{eq:4}) recovers the
usual Navier-Stokes equation for a Newtonian fluid.
This coiled state persists until the mean stretching rate is
comparable with the inverse relaxation time, $\lambda_L \tau \sim 1$
\cite{C00,BFL01}.
For larger values of $Wi$, the relaxation time of polymers is
larger than the smallest characteristic time of
the turbulent flow and molecules start to be elongated.
Figure~\ref{fig0} shows the pdf of elongations $p(R)$ at
different values of $Wi$ together with the mean square elongation.
The coil-stretch transition at $Wi^{*}=6.5 \pm 0.5$ (corresponding
to $\lambda_L \tau=0.86$) is evident.

According to the Lumley criterion \cite{L69-L73}, 
as $Wi$ increases above $Wi^{*}$, polymers start to affect the
dynamics of the turbulent cascade at the scale $\ell_{L}$
at which the eddy turnover time is of the same order of the
relaxation time, i.e. $\ell_{L} \sim (\varepsilon \tau^3)^{1/2}$.
For scales $\ell > \ell_{L}$ we expect that the turbulent cascade is
unaffected by polymers, i.e. with constant energy flux 
equal to the energy input $\varepsilon_{I}$,
while for $\ell < \ell_{L}$ we expect a reduced energy
flux $\varepsilon =\varepsilon_{\nu} < \varepsilon_I$ because 
below the Lumley scale part of the flux is removed by elastic dissipation.

\begin{figure}[htb]
\includegraphics[draft=false,scale=1.0]{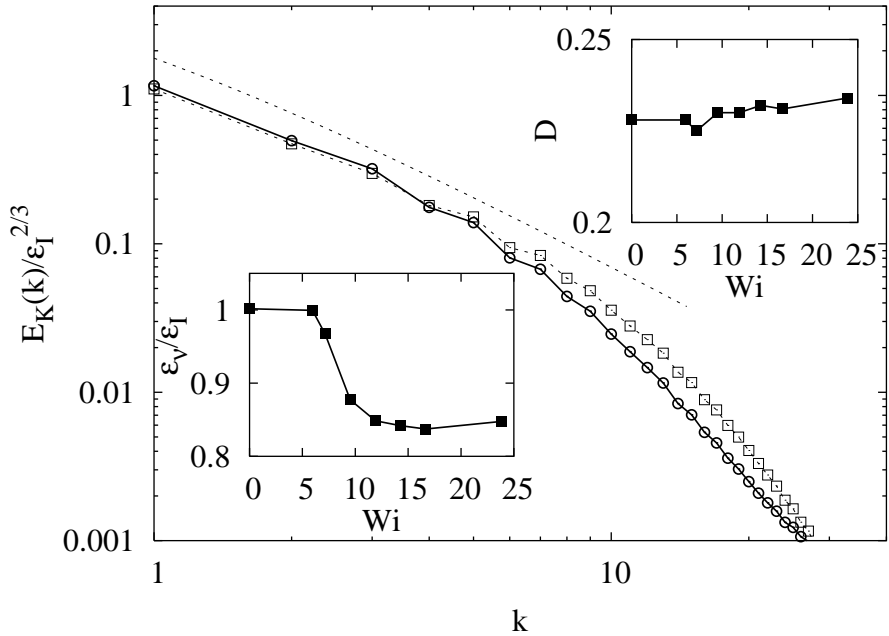}
\caption{Newtonian (squares) and viscoelastic 
($Wi=14.3$, circles) spectra of kinetic energy, normalized with 
$\varepsilon_{I}^{2/3}$. The dotted line corresponds to Kolmogorov K41 
scaling $E_{K}(k) \sim k^{-5/3}$. 
Lower inset: viscous dissipation $\varepsilon_{\nu}$ 
normalized to the energy input $\varepsilon_I$ as a function of $Wi$; the 
point size is of the order of the statistical uncertainty.
Upper inset: drag coefficient $D=\varepsilon_{I} L/E_{K}^{3/2}$
versus $Wi$.}
\label{fig1}
\end{figure}

By increasing $Wi$ above $Wi^*$ two different scenarios are possible:
the first is that elastic dissipation in (\ref{eq:5}) 
increases with $Wi$ and energy flux at scales $\ell < \ell_{L}$
vanishes (i.e. the Lumley scale $\ell_{L}$ becomes the new dissipative scale).
A second alternative is that elastic dissipation removes only
a fixed fraction of the flux. In this case $\varepsilon_{\nu}$ 
becomes independent of 
$Wi$ and thus the energy cascade proceeds below $\ell_{L}$ (with a 
reduced flux). This latter scenario has been observed in 
shell models of viscoelastic fluids \cite{BDGP03}.

Our numerical simulations at increasing values of $Wi$ 
indicate that the second scenario occurs. The inset of Fig.~\ref{fig1}
shows that the ratio $\varepsilon_{\nu}/\varepsilon_I$, which is
by definition unity for $Wi \le Wi^*$, decreases for $Wi > Wi^*$ but
already at $Wi \simeq 15 \simeq 2 Wi^*$ saturates to a new value $\sim 0.85$. 
These results are in agreement with those 
reported in \cite{LGLKT05}, where the reduction of  
vorticity in a viscoelastic solution was experimentally measured, 
and the Taylor microscale $\lambda$ was observed to be practically 
unaffected by the presence of polymers.

The above picture is supported by the comparison of the kinetic energy 
spectrum of a viscoelastic flow above the coil-stretch transition 
and the spectrum of the reference Newtonian flow.
Fig.~\ref{fig1} shows that, although the energy content at small scales is 
reduced by polymers, a power-law spectrum, characteristic of a
turbulent cascade \`a la Kolmogorov, is present in the 
viscoelastic case as well. Moreover the effect of polymers
on turbulence is {\em local} in scales, i.e. scales larger
than $\ell_{L}$ are essentially not affected by the presence of polymers.
Since the total kinetic energy $E_{K}$ is dominated by large scales,
we do not observe a significant variation of the ``drag'' coefficient,
here defined as $D=\varepsilon_{I} L/E_{K}^{3/2}$ (see the inset
in Fig.~\ref{fig1}
where it is shown that from $Wi=0$ to $Wi=24$ we measure fluctuations 
of the drag of about $2 \%$ which are within the statistical uncertainty).
This is at variance with the results reported in \cite{DCBP05}, 
where. However, it must be noticed that a different viscoelastic model was 
used, with a different large scale forcing mechanism.

\begin{figure}[htb]
\includegraphics[draft=false,scale=1.0]{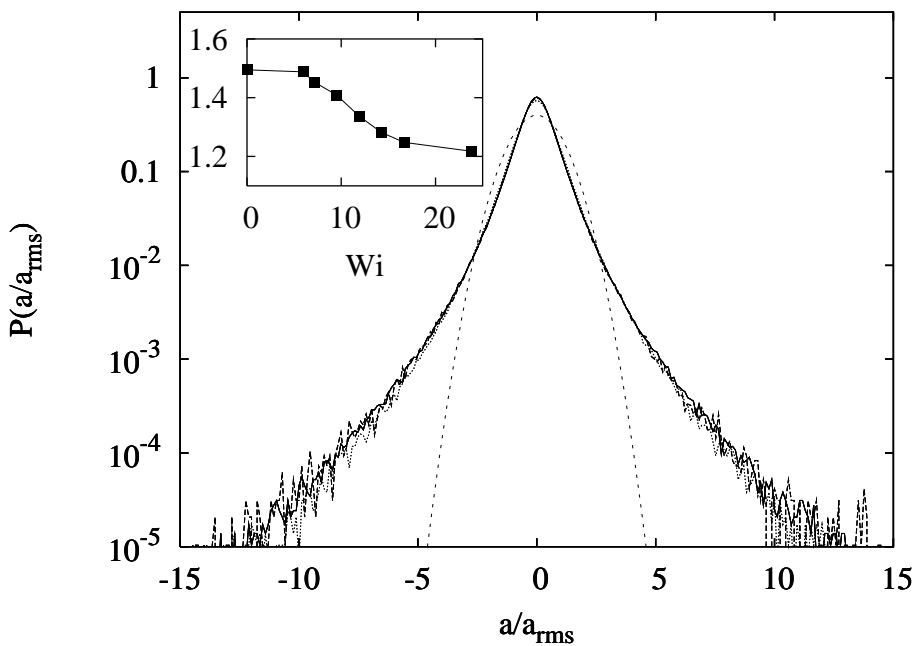}
\caption{Acceleration probability distributions 
for $Wi=0, 12, 24$;
the inner dotted curve shows a Gaussian for comparison.
Inset: total $a_{rms}$, normalized with 
$\varepsilon_{I}^{3/4} \nu^{-1/4}$, vs $Wi$.}
\label{fig2}
\end{figure}

Since the turbulent cascade survives at scales smaller than $\ell_L$,
we expect to observe some features of Newtonian turbulence in 
viscoelastic flows. One of the usual characteristics
of small scale turbulence is the extremely
intermittent acceleration which displays fluctuations much
larger than the rms value \cite{LVCAB01,BBCDLT04}.
In Figure~\ref{fig2} we report the statistics for the acceleration
in viscoelastic turbulent flows at different values of elasticity
above $Wi^*$ compared with the Newtonian case. 
In presence of polymers, the rms value $a_{rms}$ is 
reduced with respect to the Newtonian case, again in agreement with 
experimental observations \cite{CMLB02}. For sufficiently
large values of $Wi$, $a_{rms}$ becomes almost independent of $Wi$
(and, at the largest value $Wi=24$, is about $80 \%$ of the Newtonian case).
This is consistent with a reduced value of energy flux at small
scales shown in Fig.~\ref{fig1}: indeed by compensating $a_{rms}$ with 
the dimensional estimation
$\varepsilon_{\nu}^{3/4} \nu^{-1/4}$ this becomes
virtually independent of $Wi$.

The probability density functions shown in Fig.~\ref{fig2} indicate
that relative fluctuations of turbulent acceleration are
not affected by the presence of polymers.
Indeed, the pdf of the rescaled quantity $a/a_{rms}$ is found to be 
$Wi$-~independent in all the range of $Wi$ investigated. 
This is a remarkable result which, besides its intrinsic interest,
has an important practical consequence: the fact that
the {\em shape} of the acceleration pdf is not affected by polymers
implies that it is possible to model small scale statistics in viscoelastic 
turbulence below the Lumley scale by means of the same models used 
for Newtonian fluids, by simply changing global quantities such
as the energy flux.

\begin{figure}[hbt]
\includegraphics[draft=false,scale=1.0]{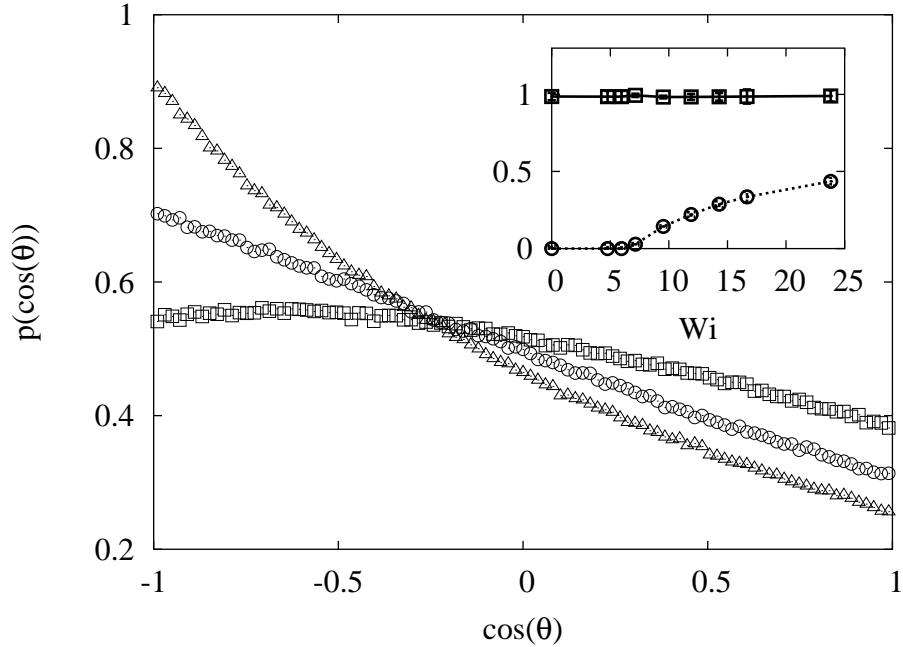}
\caption{Probability density functions 
of the cosine of the angle $\theta$ between the vectors 
$-{\bm \nabla} p$ and ${\bm R} \cdot {\bm \nabla} {\bm R}$
for $Wi=9.5$ (squares), $Wi=14.3$ (circles) and $Wi=24.0$ (triangles).
Inset: Leading contributions to the acceleration, 
normalized with the total rms acceleration,
as a function of $Wi$: 
pressure gradient $a_p/a_{rms}$ (squares)
and elastic stress contribution $a_{el}/a_{rms}$ (circles).}
\label{fig3}
\end{figure}

The acceleration ${\bm a}=d{\bm u}/dt$ in a viscoelastic flow 
has three different contributions from the rhs of (\ref{eq:4}): pressure
gradients, viscous and elastic contributions. In fully developed
turbulence the viscous contribution is negligible and one may
ask which is the dominant contribution for $Wi>Wi^*$.
Figure~\ref{fig3} (inset) shows that the pressure gradient contribution
$a_{p}=\langle ({\bm \nabla} p)^2\rangle^{1/2}$
(which is the only term in the Newtonian limit $Wi=0$)
is always dominant in the range of $Wi$ investigated. 
Nevertheless, the contribution of polymers is not negligible: at
the largest $Wi=24$ the rms value of the elastic acceleration
$a_{el}= (2 \nu /\tau)\langle ({\bm R} \cdot {\bm \nabla} 
{\bm R})^2\rangle^{1/2}$
is almost $50 \%$ of the total acceleration $a_{rms}$.

Since ${\bm a} \simeq -{\bm \nabla} p + 
(2 \nu /\tau) {\bm R} \cdot {\bm \nabla} {\bm R}$, 
the fact that $a_{p} \simeq a_{rms}$
means that, increasing $Wi$, the flow develops strong correlations
between the pressure gradient and the elastic component in (\ref{eq:4}).
As a measure of this correlation we have computed the angle $\theta$
between the vectors $-{\bm \nabla} p$ and ${\bm R} \cdot {\bm \nabla} {\bm R}$.
Figure~\ref{fig3} displays the pdf of $cos(\theta)$ and shows that, 
indeed, increasing $Wi$ the two vectors tend to be anticorrelated
(i.e. $cos(\theta)=-1$) with higher and higher probability.

In conclusion, we have studied the statistics of turbulent 
polymer solutions within the uniaxial model of viscoelastic flow.
We have found numerical evidence for a coil-stretch transition
at $Wi^*$ above which polymers affect the small scales of the 
turbulent flow.
For $Wi>Wi^*$, the energy flux is partially removed by 
polymer elasticity at the Lumley scale. This effect
saturates and the turbulent cascade proceeds to scales
smaller than the Lumley scale.
As a consequence, small scale statistics, such as acceleration, 
display features typical of Newtonian turbulence.

\acknowledgments
We acknowledge the support of MIUR-Cofin 2005
``Meccanica statistica dei sistemi complessi''.
Numerical simulations have been performed on the
``Turbofarm'' cluster at the INFN computing center in Torino.


\end{document}